\documentclass[aps,prb,reprint,superscriptaddress]{revtex4-2}
\usepackage{amsmath}
\usepackage{amssymb}
\usepackage{graphicx}
\begin{document}
\title{Magneto-optical Kerr effect and magnetoelasticity in weak ferromagnetic RuF$_4$ monolayer}
\author{Na Wang}
\author{Jun Chen}
\author{Ning Ding}
\affiliation{School of Physics, Southeast University, Nanjing 21189, China}
\author{Huimin Zhang}
\affiliation{Key Laboratory of Computational Physical Sciences (Ministry of Education), Institute of Computational Physical Sciences, and Department of Physics, Fudan University, Shanghai 200433, China}
\affiliation{Shanghai Qi Zhi Institute, Shanghai 200030, China}
\author{Shuai Dong}
\email{Email: sdong@seu.edu.cn}
\author{Shan-Shan Wang}
\email{Email: wangss@seu.edu.cn}
\affiliation{School of Physics, Southeast University, Nanjing 21189, China}

\date{\today}

\begin{abstract}
Considerable research interest has been attracted to noncollinear magnetic structures for their intriguing physics and promising applications. In this work, based on relativistic density functional theory, we reveal the interesting magnetic order and relevant properties in monolayer RuF$_4$, which can be exfoliated from its bulk phase. Although the spins on Ru ions are almost antiferromagnetically aligned between nearest-neighbors, weak ferromagnetism is generated because of the antisymmetric Dzyaloshinskii-Moriya interaction as well as the single-ion anisotropy. A prominent magneto-optical Kerr effect can be observed for this antiferromagnet, similar to those of regular strong ferromagnets. In addition, a uniaxial strain can induce a ferroelastic switching together with the in-plane rotation of spin direction, giving rise to a strong intrinsic magnetoelasticity. Our work not only suggests an alternative direction for two-dimensional magnetic materials, but also provides hints to future devices based on antiferromagnetic magnetoelastic or magneto-optical materials.
\end{abstract}
\maketitle

\section{Introduction}
Two-dimensional (2D) magnetic materials have attracted many research attentions, which have promising prospects in the miniaturization of spintronic devices. Recently, a few 2D magnets have been realized in experiments, involving CrI$_{3}$, VSe$_{2}$, and so on \cite{CrI3,VSe2, MnSe2,Fe3GeTe2}. So far, most of these pioneering works on 2D magnets are focused on $3d$ transition metal compounds with ferromagnetism \cite{An:APL}, leaving the 2D $4d$/$5d$ antiferromagnets rarely touched. Those $4d$/$5d$ metal compounds with partially filled shells usually demonstrate the combination of strong spin-orbit coupling (SOC) effects and non-negligible correlation effects, making them a wonderful platform to explore a wide variety of spin textures and relevant non-trivial physical properties.

Current mainstream spintronic devices are based on ferromagnetic (FM) materials. Until recently, the concept of antiferromagnetic (AFM) devices has re-shaped the future of spintronics \cite{Wunderlich2016, Tserkovnyak2018, TserkovnyakNP2018, KimelNP2018, Ohno2018}. There are at least two important advantages for AFM spintronics. First, they have better scalability without the influence from stray field and are more robust against magnetic field perturbations \cite{Tserkovnyak2018}. Second, the resonance frequencies in antiferromagnets are typically improved to THz range, which can lead to much higher operation speeds compared to those ferromagnetic counterparts \cite{Felser2018}.

However, there remain many challenges in those devices based on pure antiferromagnets. Unlike ferromagnets with net magnetic moments that can be manipulated by external magnetic field, the absence of net magnetization in antiferromagnets makes them much less sensitive to external magnetic field and thus hard to be manipulated. Hence, an ideal system with mainly antiferromagnetic textures and net ferromagnetic moments may take full advantage of both sides. Such appealing systems have been proposed and detected in three-dimensional materials, such as antiferromagnetic FeBO$_3$, $R$FeO$_3$ ($R$: rare earth), Sr$_2$IrO$_4$, BiFeO$_3$, and other bulks \cite{DMIFeBO3EXP,RFeO3,BiFeO3dft,Sr2IrO4Dong,wang2019giant}. The weak ferromagnetism in these materials is caused by antisymmetric interaction or the single-ion anisotropy (SIA). Furthermore, anomalous Hall effect and Kerr effect have been studied in Mn$_3X$ ($X$= Sn, Ir, Pt) and RuO$_{2}$ with weak ferromagnetism \cite{AHEMn3Sn,PhysRevB.92.144426,PhysRevB.104.024401,higo2018large}. However, such weak ferromagnetism has not been widely concerned in 2D magnetic materials.

In addition, many 2D materials have been proposed to possess ferroelasticity. The 1T'-transition metal dichalcogenides and doped GdI$_3$ monolayers have three equivalent orientation variants which are switchable under mechanical strain \cite{WTe2fEL,YouPrb}. $\alpha$-ZrPI was predicted to be ferroelastic with anisotropic carrier mobility \cite{MPIFEL}. In 1T'-MoTe$_2$ and $\beta$'-In$_2$Se$_3$, ferroelasticity has been observed in experiments \cite{MoTe2,In2Se3}. Particularly, ferroelasticity coupling with magnetism, denoted as magnetoelasticity, can expand the potential applications in spintronic devices. Only a few researches have reported intrinsic magnetoelasticity in 2D materials. For example, monolayer FeOOH, VF$_4$, and VNI were predicted to display magnetoelasticity \cite{FeOOH,VF4,VNI}. Hence, exploring more candidates with intrinsic strong magnetoelasticity coupling is crucial in 2D materials.

In this work, we demonstrate that monolayer RuF$_4$ can be exfoliated from its bulk phase with little energy cost. The partially filled Ru's $4d$ shells provide magnetic moments. It is an intrinsic antiferromagnetic semiconductor with weak ferromagnetism from the spin canting, driven by the Dzyaloshinskii-Moriya (DM) interaction and SIA. The Monte Carlo (MC) simulation confirms its noncollinear antiferromagnetism and weak ferromagnetism below $T_{\rm N}=45.5$ K. Remarkably, a longitudinal magneto-optical Kerr effect (MOKE) is expected due to the spatial inversion  combined with time-reversal ($PT$) symmetry breaking. Another interesting property is the intrinsic coupling of antiferromagnetism and ferroelasticity. Its spin axis can be manipulated by the reversible uniaxial strain, leading to a strong intrinsic magnetoelasticity.

\section{Methods}
Our first-principles calculations were based on the density functional theory (DFT), as performed using the Vienna \emph{ab-initio} Simulation Package (VASP) \cite{kresse1993ab,kresse1996g}. The interactions of ions and electrons were treated by the projector augmented wave method \cite{blochl1994pe}. The generalized gradient approximation (GGA) parameterized by Perdew, Burke, and Ernzerhof (PBE) was adopted to describe the exchange-correlation functional \cite{perdew1996generalized}. %To simulate the correlation effect associated with the Ru-$4d$ orbitals, the Hubbard $U$ correction implemented in the PBE$+U$ method was included in our calculations \cite{dudarev1998electron}, with $U$=$x$ eV \cite{xx}.

For the bulk calculation, the van der Waals (vdW) D3 correction is adopted \cite{VDWD3}. For monolayer calculation, a vacuum space of $20$ {\AA} thickness was introduced to avoid layer interactions. The $\Gamma$-centred $k$-mesh with $11\times11\times1$ was adopted for the Brillouin zone sampling \cite{monkhorst1976special}. The energy convergence criterion was set to be $10^{-6}$ eV. The lattice parameters and ionic positions were fully optimized until the residual force on each atom was less than $0.001$ eV/{\AA}. The phonon spectrum was calculated with DFPT method using PHONOPY package \cite{Phonopy}.

To calculate the MOKE signal, the \emph{exciting} code was used \cite{excitingcode}, with the (linearized) augmented plane waves plus a local orbital basis \cite{MOKEmethod}. The Muffin-Tin radii for Ru and F atoms are $2.0$ and $1.45$ Bohr, respectively. SOC is included using the second-variation method. The $k$-point grids of $5\times5\times1$ were sampled in the first Brillouin zone. The \emph{exciting} code uses the time-dependent density-functional theory (TDDFT) to calculate the Kerr angle ($\theta_K$) and ellipticity ($\eta_{K}$) \cite{es,pccp,ALKAUSKAS20101081}.

In addition, the magnetic transition was simulated using the Markov-chain MC method with Metropolis algorithm \cite{MC}. Metropolis algorithm combined with parallel tempering process was adopted to obtain the thermal equilibrium efficiently \cite{Nemoto1996}. A $48\times48$ lattice with periodic boundary conditions was used in the MC simulations to avoid nonuniversal effects of boundary conditions. We took an exchange sampling after every $10$ standard MC steps. The initial $4\times10^5$ MC steps were discarded for thermal equilibrium and another $4\times10^5$ MC steps were retained for statistical average in the simulation. The specific heat $C$  was also calculated to determine $T_{\rm N}$. The results were further checked by the simulations on lattices with other sizes to ensure that the finite-size effect would not affect our conclusion.

\section{Results \& Discussion}
\subsection{Structure \& Stability}

\begin{figure}[t]
\centering
\includegraphics[width=0.48\textwidth]{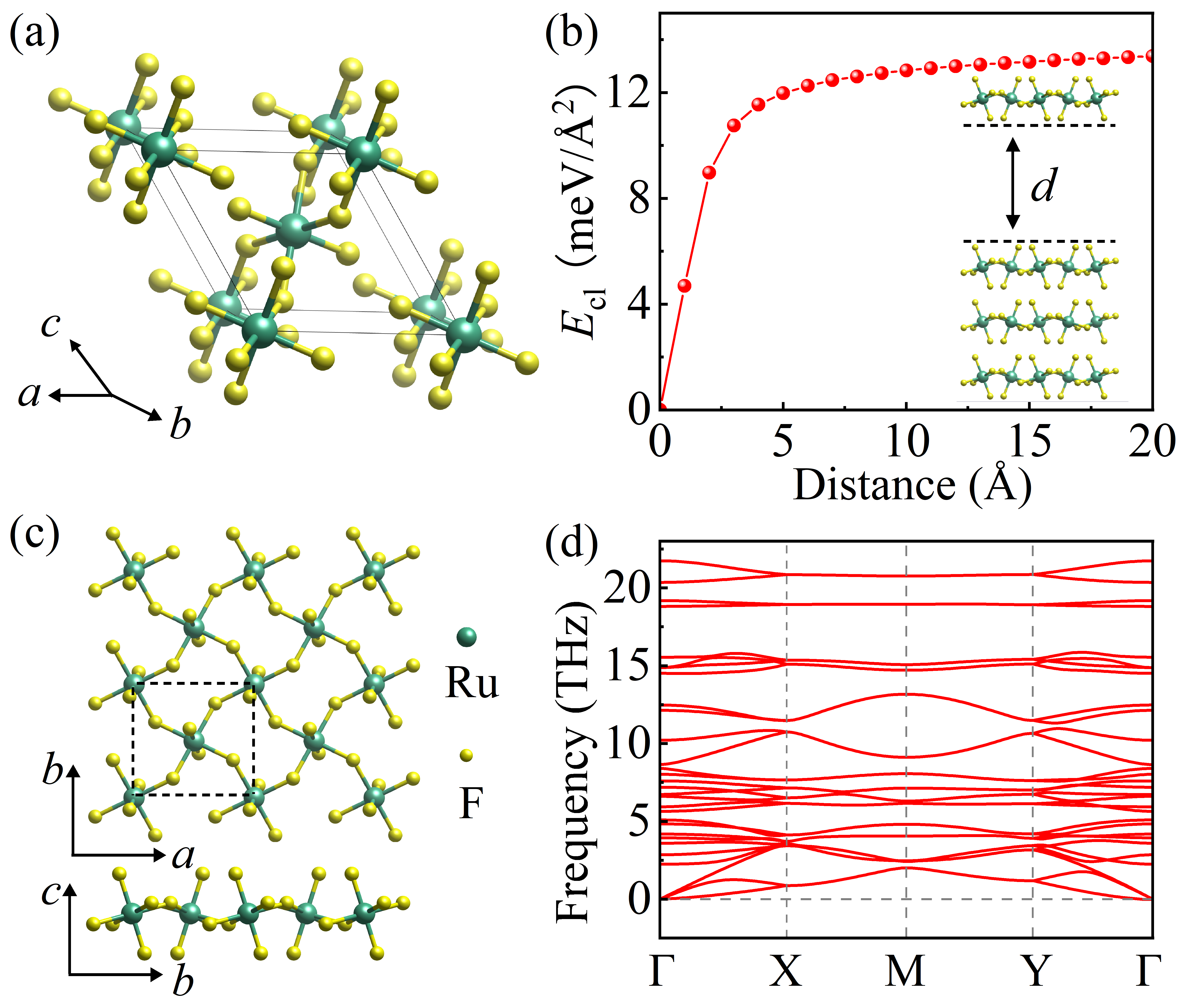}
\caption{Crystal structure of RuF$_4$. (a) Bulk phase of RuF$_4$. The vdW layers stack along the [101] direction. (b) Schematic of exfoliation process. (c) Top and side views of monolayer RuF$_4$. The dashed rectangle  indicates the primitive unit cell with ten atoms. (d) Phonon spectrum of monolayer RuF$_4$. A $4\times4$ supercell is used in the calculation.}
\label{fig1}
\end{figure}

Bulk phase ruthenium tetrafluoride was experimentally synthesized in 1992 \cite{RuF4experiment}. It has a layered monoclinic structure consisting of ten atoms per unit cell with space group $P2_1/c$ (No. 14). Each ruthenium ion is surrounded by six fluorine ions, forming an octahedron. And neighboring RuF$_6$ octahedra are connected via the corner-sharing manner. The experimental lattice constants are $a=5.607$ {\AA}, $b=4.946$ {\AA}, and $c=5.413$ {\AA} with RuF$_4$ layers stacking along the [101] direction, as shown in Fig.~\ref{fig1}(a). The interlayer coupling is vdW interaction, implying that single layer RuF$_4$ could be exfoliated  from its bulk phase easily. Our calculated lattice constants ($a=5.610$ {\AA}, $b=5.054$ {\AA}, and $c=5.532$ {\AA}) are consistent with the experimental results, providing a solid starting point for the following calculations.

The cleavage energy is evaluated by calculating the energy variation of a single layer exfoliated from its bulk phase, as shown in Fig.~\ref{fig1}(b), which is $14$ meV/\AA$^2$, smaller than those of many familiar 2D materials such as graphene ($\sim21$ meV/\AA$^2$), MoS$_2$ ($\sim18$ meV/\AA$^2$), and black phosphorene ($\sim23$ meV/\AA$^2$) \cite{Exfoliation}. Therefore, it is hopeful to obtain monolayer RuF$_4$ from its bulk phase using mechanical approaches in experiments.

The top and side views of monolayer RuF$_4$ are depicted in Fig.~\ref{fig1}(c). The monolayer displays a tetragonal lattice with the same space group  $P2_1/c$  containing $4$ symmetry elements. The fully optimized lattice constants of monolayer RuF$_4$ are $a=5.436$ \AA\ and $b=5.125$ \AA. The phonon spectrum is evaluated to verify its dynamical stability. As shown in Fig.~\ref{fig1}(d), there is no imaginary phonon branch through the whole Brillouin zone, demonstrating the monolayer structure is dynamically stable.

\subsection{Weak ferromagnetism}
\begin{figure}
\centering
\includegraphics[width=0.48\textwidth]{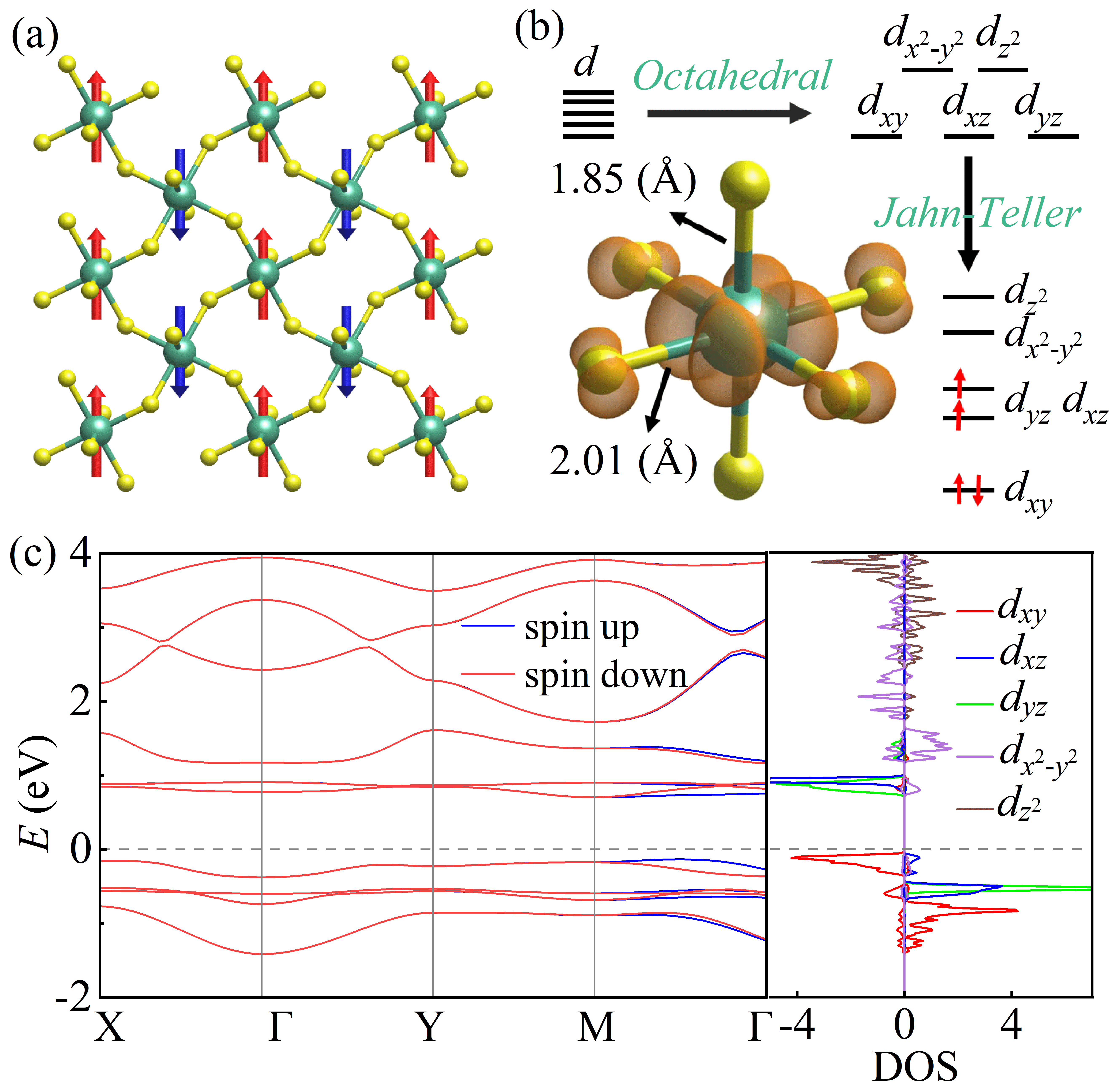}
\caption{Electronic structure of monolayer RuF$_4$. (a) N\'eel AFM configuration, which is the most energetically favored magnetic order. (b) Electron configuration of Ru's $4d$ orbitals, as a result of octahedral crystal field and Jahn-Teller distortion. (c) Band structure and orbital-projected DOS.}
\label{fig2}
\end{figure}

The unpaired electrons in $4d$ orbitals usually induce magnetic moments in transition metal compounds, which is exactly the case for monolayer RuF$_4$. The ground state of monolayer RuF$_4$ is determined by comparing the energy among four magnetic configurations using a $2\times2$ supercell: FM, double stripe-antiferromagnetic (AFM-I), N\'eel-antiferromagnetic (AFM-II), and stripe-antiferromagnetic (AFM-III), as shown in Fig.~S1 of Supplementary Materials (SM) \cite{supp}. It is found that the N\'eel-antiferromagnetic state is the most energetically favorable one, as depicted in Fig.~\ref{fig2}(a). The energies of FM, AFM-I, and AFM-III states are $14.25$, $5.85$, and $5.80$ meV/Ru higher than the ground state. In the AFM-II state, the magnetic moments are mainly distributed on Ru sites with $\sim1.45$ $\mu_{\rm B}$/Ru.

Due to the octahedral crystal field, the $d$ orbitals will be split into two groups: the low-lying $t_{\rm 2g}$ triplets and higher-energy $e_{\rm g}$ doublets. The compressed Jahn-Teller distortion ($Q_3$ mode) of RuF$_6$ octahedron (i.e., two out-of-plane Ru-F bonds $1.85$ {\AA} are shorter than the four in-plane ones $2.01$ \AA) further splits the $t_{\rm 2g}$ orbitals. The on-site energy of $d_{xy}$ orbitals is lower than that of degenerate $d_{yz}$ and $d_{xz}$ orbitals. In the case of Ru$^{4+}$ with four $4d$ electrons, three spin-up electrons occupy the three $t_{\rm 2g}$ orbitals and one more electron is placed in the low-lying spin-down $d_{xy}$ orbital, as shown in Fig.~\ref{fig2}(b). This orbital configuration leads to a local magnetic moment up to $2$ $\mu_{\rm B}$, close to the DFT one. Such an orbital configuration also quenches most orbital moments \cite{Weng:Prb}, as confirmed in our DFT calculation ($0.001$ $\mu_{\rm B}$/Ru from the orbital contribution).

The electronic structure of AFM-II is plotted in Fig.~\ref{fig2}(c). The band structure exhibits a semiconducting behavior with a direct band gap around $0.86$ eV. In the orbital-projected density of states (DOS), the $d_{xy}$/$d_{xz}$/$d_{yz}$ spin-up orbitals and $d_{xy}$ spin-down orbital locate below the Fermi level, while the $d_{xz}$/$d_{yz}$ spin-down orbitals are higher than Fermi level. This result is consistent with the above analysis of electron configuration.

\begin{figure*}
\includegraphics[width=0.96\textwidth]{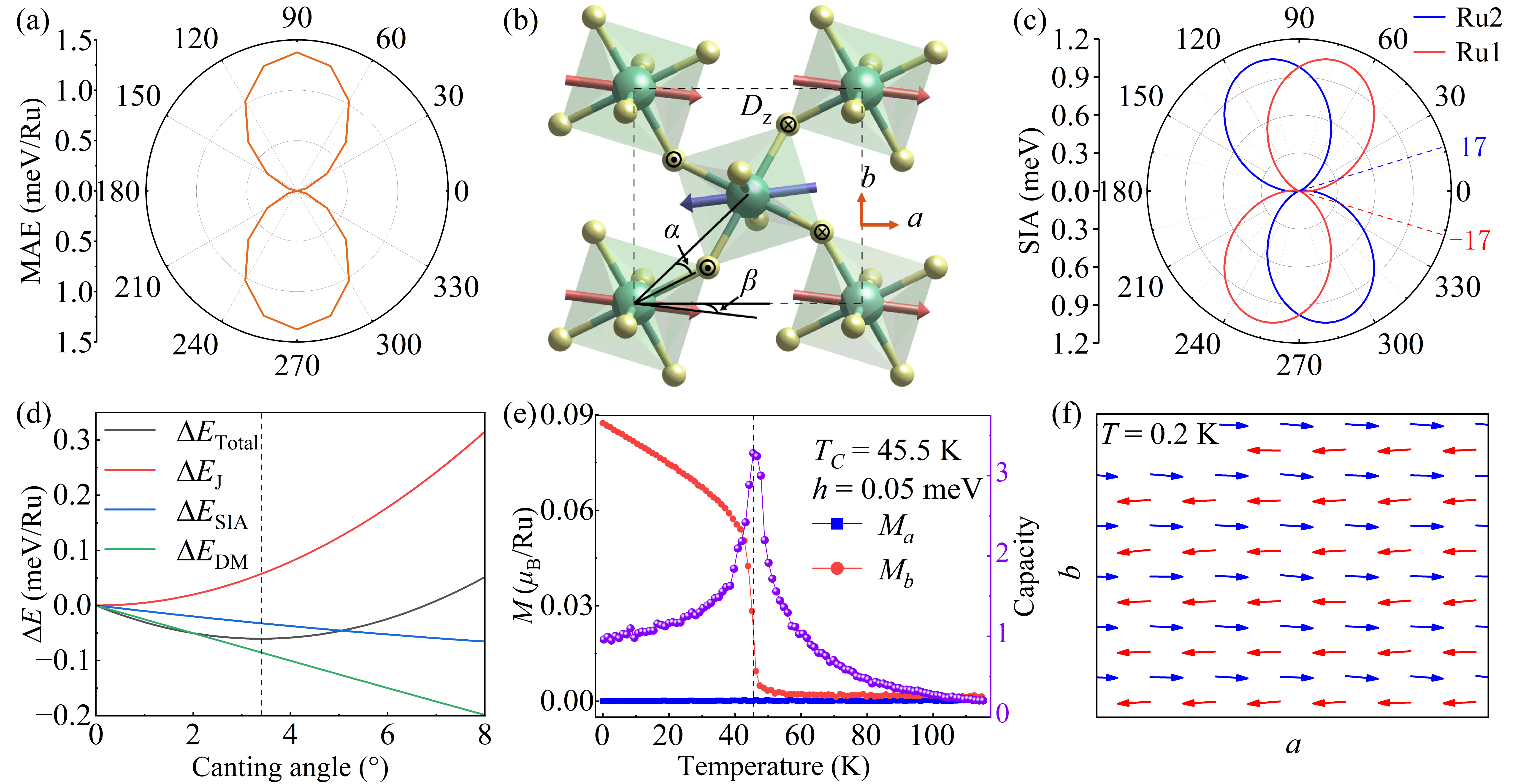}
\caption{Magnetism of monolayer RuF$_4$. (a) MAE as a function of global spin orientation. It is clear that the easy axis is along the $a$-axis. (b) Spin canting of AFM. $\alpha$ represents the octahedral rotation angle and $\beta$ denotes the spin canting angle. The $z$-components of $\textbf{D}$ vectors for Ru-F-Ru bonds are indicated by arrows ($\bigodot$ and $\bigotimes$). (c) The SIA energy profile for two Ru sites. (d) The energy contributions from each term as a function of canting angle $\beta$. (e) The MC simulated heat capacity $C$ and two components of magnetization as a function of temperature. (f) The MC snapshot of a canting spin pattern at low $0.2$ K. A small biased magnetic field ($0.05$ meV) is applied along the $b$-axis during the MC simulation to polarize the net magnetization.}
\label{fig3}
\end{figure*}

In AFM RuF$_4$, the RuF$_6$ octahedra are complemented by the non-magnetic F atoms. The octahedra are distorted with collective rotation and tilting, which results in the staggered displacements of in-plane F ions, as indicated in Fig.~\ref{fig2}(a). The combination of these distorted octahedra (or in other word the staggered F ions) and the AFM-II ordering breaks the $PT$ and $t_{1/2}T$ symmetries (where $t_{1/2}T$ is a half unit cell translation), which then allows weak ferromagnetism \cite{Dong:Prl2009}. To account for this, the preferred spin orientation is determined by calculating the global magnetocrystalline anisotropy energy (MAE). First, the out-of-plane direction is found to be a hard axis with a strong MAE ($7.23$ meV/Ru higher), as a result of aforementioned orbital configuration.  Second, the in-plane MAE shows that the favorable magnetic orientation of monolayer RuF$_4$ is along the $a$-axis, as shown in Fig.~\ref{fig3}(a). Then its magnetic point group becomes $2/m$, containing a mirror plane and two-fold axis. With this low symmetric magnetic point group, a spin canting to the $b$-axis is allowed. To confirm the spin canting and estimate its magnitude, the magnetic structure is relaxed with SOC, starting from initial collinear AFM spins along the $a$-axis. The SOC effects indeed lead to a noncollinear spin texture. Specifically, the spins of Ru ions tilt from the $a$-axis to $b$-axis by $\pm3.4^\circ$, as shown in Fig.~\ref{fig3}(b). Such spin tilting angles result in a weak but measurable FM moment $\sim0.11$ $\mu_{\rm B}$/Ru along the $b$-axis.

Although the above symmetry analysis and DFT calculation confirm the weak ferromagnetism due to spin canting, it is essential to clarify its microscopic origin. In general, a small spin canting angle can be driven by antisymmetric DM interaction \cite{Dzyaloshinsky1958,Moriya1960}, or site-dependent SIA, both of which are SOC effects. The strong SOC of $4d$ orbitals and the staggeringly bending Ru-F-Ru bonds can lead to staggered DM vectors \cite{Dong:Prl2009}, as shown in Fig.~\ref{fig3}(b). Also, the collective rotation of RuF$_6$ octahedra can give rise to site-dependent SIA, which is beyond the above scenario of global magnetocrystalline anisotropy. Then a $XY$-spin model can be constructed to describe its magnetism, while the high-energy out-of-plane spin component is neglected. The model Hamiltonian reads as:
\begin{equation}
H=\sum_{\langle i,j\rangle}[J({\textbf S}_i\cdot{\textbf S}_j)+{\textbf D}_{ij}\cdot({\textbf S}_i\times{\textbf S}_j)]+\sum_iK_{ab}({\textbf S}_i\cdot{\textbf u}_i)^2,
\label{eq1}
\end{equation}
where the first and second terms are the normal super-exchange and the DM interaction between nearest-neighboring (NN) sites $i$ and $j$; and the third term is the in-plane SIA. ${\textbf S}$ represents the normalized spin and ${\textbf u}_i$ is the normalized vector along the in-plane easy axis at site $i$. The DFT extracted magnetic coefficients are listed in Table~\ref{table1}. The obtained $\textbf{u}_i$'s are close to the global easy axis but with $\pm17^\circ$ canting, as shown in Fig.~\ref{fig3}(c).

\begin{table}
\caption{Magnetic coefficients (in unit of meV) for the spin model, extracted from DFT calculations. For comparison, the out-of-plane MAE coefficient $K_c$ is also shown, which is positive and large enough to make all spins lie in-plane. Thus, only the $z$-component of $\textbf{D}$ ($D^z$) is valuable. The SIA axis for two Ru sites are also shown.}
\centering
\begin{tabular*}{0.48\textwidth}{@{\extracolsep{\fill}}cccccc}
\hline \hline
$J$ & $D^z$ & $K_{ab}$ & $K_c$ & $\textbf{u}_1$ & $\textbf{u}_2$\\
\hline
$4.07$ & $-0.36$ & $-1.07$ & $7.23$ & ($0.956$, $0.293$) & ($0.956$, $-0.293$)\\
\hline \hline
\end{tabular*}
\label{table1}
\end{table}

With this spin model, we can clarify the origin of spin canting. The energy contributions from each term as a function of canting angle are plotted in Fig.~\ref{fig3}(d). Although the coefficient $K_{ab}$ is larger than $D^z$, the DM interaction contributes more to the energy change than the SIA one. If only $J$ and SIA are considered, the total energy reaches its minimum with a canting angle of $1^\circ$. If only $J$ and DM interaction are considered, the canting angle reaches $2.5^\circ$, closer to the final value ($3.4^\circ$). Hence, the DM interaction plays the dominant role in the formation of weak FM, although the SIA also helps.

Having all these coefficients, the magnetic phase transition of monolayer RuF$_4$ is simulated using the MC method, as shown in Fig.~\ref{fig3}(e). The N\'eel temperature $T_{\rm N}$ is estimated as $45.5$ K, indicated by the peak of heat capacity. The MC snapshot far below $T_{\rm N}$ also confirms its canting antiferromagnetism and weak ferromagnetism, as shown in Fig.~\ref{fig3}(f).

\subsection{Magneto-optical Kerr Effect}
To characterize the ferromagnetism in 2D materials, MOKE is the most frequently used method. The presence of MOKE depends on the breaking of $PT$ symmetry. Microscopically, a nonzero MOKE signal requires the SOC effects and net magnetization \cite{MOKE}. According to the directional relationship between the magnetization, the reflecting surface, and the plane of incidence, there are three kinds of MOKE. The polar MOKE (P-MOKE) is the most common case, with the magnetization perpendicular to the reflection surface and parallel to the plane of incidence. For the longitudinal MOKE (L-MOKE), the magnetization is parallel to both the reflection surface and the plane of incidence, while in the case of transversal MOKE (T-MOKE) the magnetization is perpendicular to the plane of incidence and parallel to the surface.

Since the net magnetization of RuF$_4$ monolayer lies in-plane, its MOKE effect belongs to L-MOKE. Required by the  symmetric operations of magnetic point group $2/m$, the general form of monolayer RuF$_4$'s dielectric tensor can be expressed as \cite{born2013principles}:
\begin{equation}
\varepsilon=\left[\begin{array}{ccc}
    \varepsilon_{xx} & 0 & \varepsilon_{xz} \\
    0 &  \varepsilon_{yy} & 0 \\
    \varepsilon_{zx} & 0 & \varepsilon_{zz}
    \end{array}\right],
\label{DE}
\end{equation}
which has five independent coefficients. Since the dielectric tensor is associated with optical conductivity via $\varepsilon_{ij}(\omega)=\delta_{ij}+i\frac{4\pi}{\omega}\sigma_{ij}(\omega)$ \cite{MOKEtensor,MOKEtensor2}, the optical conductivity of monolayer RuF$_4$ can be expressed as:
\begin{equation}
\sigma=\left[\begin{array}{ccc}
    \sigma_{xx} & 0 & \sigma_{xz} \\
    0 &  \sigma_{yy} & 0 \\
    \sigma_{zx} & 0 & \sigma_{zz}
    \end{array}\right].
\label{OE}
\end{equation}

\begin{figure}
\includegraphics[width=0.48\textwidth]{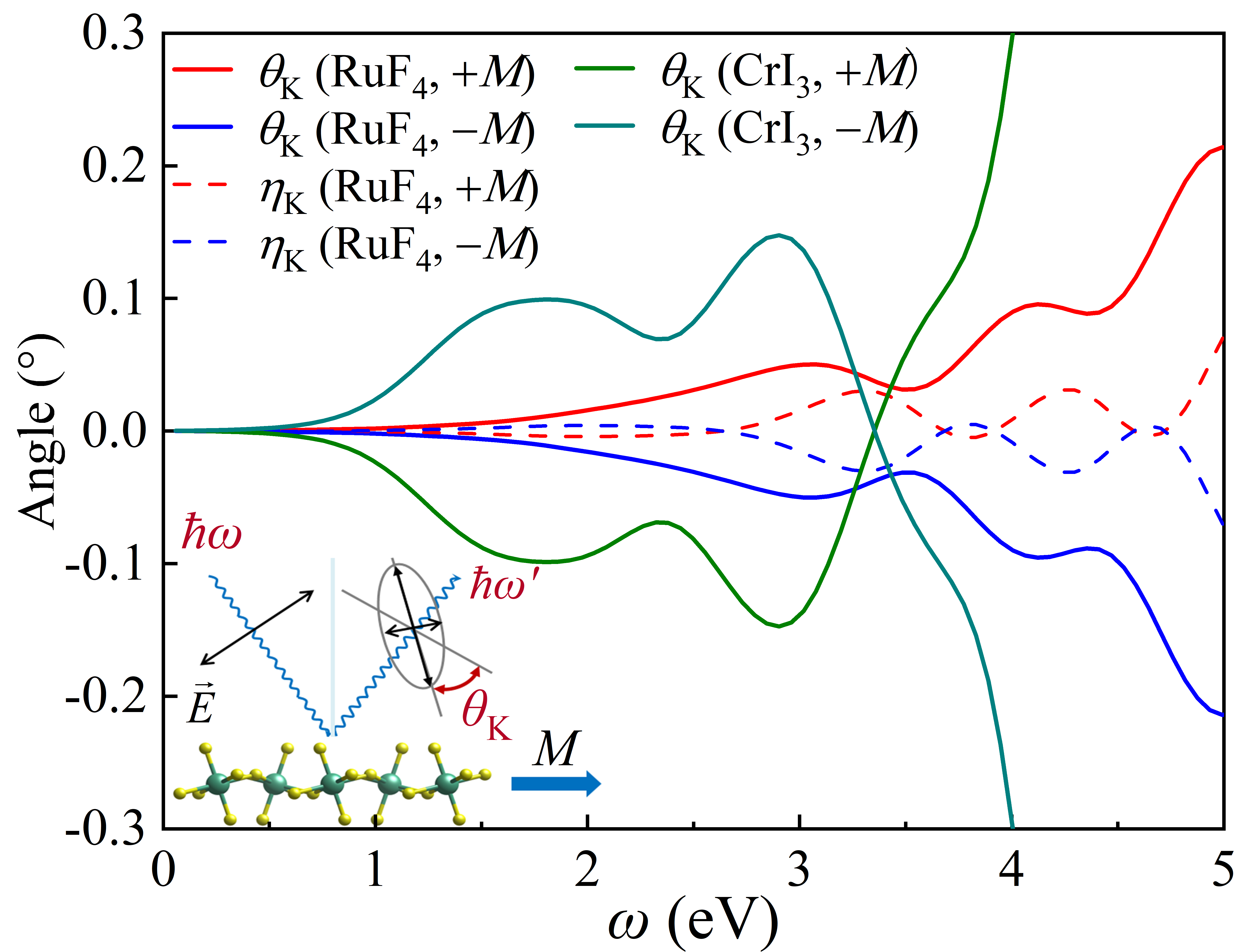}
\caption{Calculated L-MOKE signals of monolayer RuF$_4$. For comparison, the P-MOKE Kerr rotation angle of pure FM monolayer CrI$_3$. Insert: schematic of MOKE measurement. For L-MOKE (P-MOKE), the incident light should be nearly grazing on (normal to) the 2D plane with photon energy $\hbar\omega$. The polarization state changes for reflected light, leading to rotation ellipticity.}
\label{fig4}
\end{figure}

For L-MOKE, the complex Kerr angle should be calculated using the equation:
\begin{equation}
\phi_K=\theta_K+i\eta_K=\frac{-(\sigma_{xz}-\sigma_{zx})}{2\sigma_{xx}\sqrt{1+\frac{i4\pi}{\omega}\sigma_{xx}}},
\label{3}
\end{equation}
where the real part $\theta_K$ corresponds to the Kerr rotation angle, and the imaginary part $\eta_K$ refers to the Kerr ellipticity. The calculated L-MOKE signal is shown as Fig.~\ref{fig4}, which is odd to magnetization as expected.

Here the value of Kerr angle is slightly smaller than that of monolayer CrI$_3$ (which is the P-MOKE) in most energy window \cite{CrI3MOKE}. Even though, it remains significant, especially considering its very small magnetization (only $3\%$ of CrI$_3$), which is due to the stronger SOC of $4d$ orbitals than $3d$ ones.

\subsection{Strong magnetoelasticity}
\begin{figure}
\includegraphics [width=0.48\textwidth]{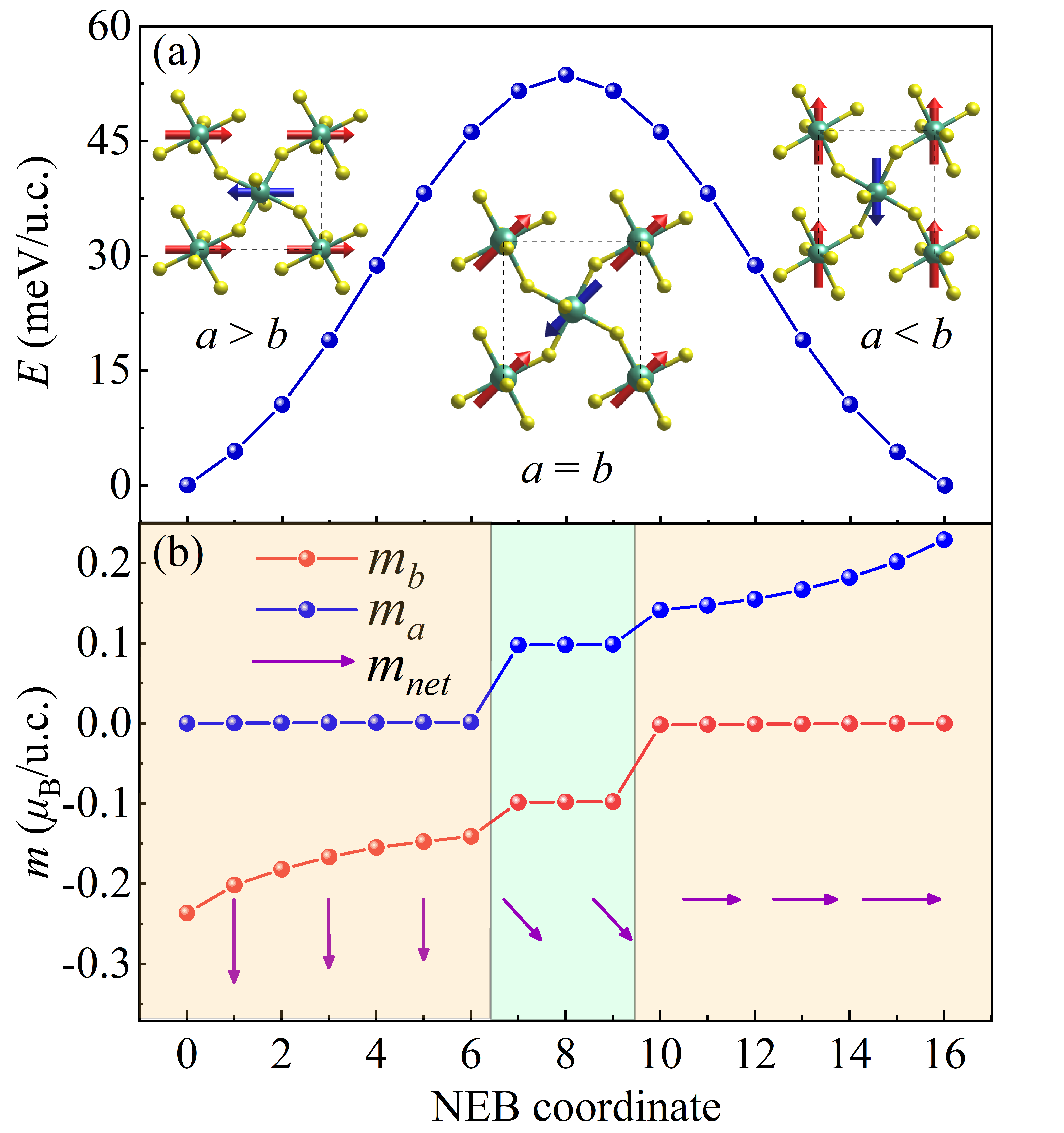}
\caption{(a) The energy barrier of the ferroelastic switching calculated by CI-NEB method. (b) The magnitude and direction of net magnetization during the ferroelastic switching. }
\label{fig5}
\end{figure}

In addition to the nontrivial magnetic-optical property, the monolayer RuF$_4$ also exhibits strong ferroelasticity. Ferroelastic materials have two or more equally stable states that can be interchanged from one to another variant, driven by external strains \cite{Khomskii2009,FELreview,YouPrb}. Here the two ferroelastic states are shown in Fig.~\ref{fig5}(a): the $S$ state indicates lattice constant $a$ larger than $b$, while $S'$ state can be understood as the $90^\circ$-rotated variant of $S$ state. The ferroelastic switching from $S$ state to $S'$ state can be realized by applying a uniaxial strain, as shown in Fig.~\ref{fig5}(a). The paraelastic state $P$ can be defined as the intermediate one between the two ferroelastic states, with $a=b=5.372$ \AA.

The ferroelastic switching process is simulated using the climbing image-nudged elastic band (CI-NEB) method \cite{NEB}. The calculated transition barrier is $53.6$ meV/u.c. ($5.36$ meV/atom) [Fig.~\ref{fig5}(a)], which is much smaller than that of phosphorene ($\sim200$ meV/atom) \cite{Wu2016}, and borophene ($\sim100$ meV/atom) \cite{FELborophene}.

Another key factor of ferroelasticity is the ferroelastic strain, which can be described by a $2\times2$ transformation strain matrix ($\eta$) from the Green-Lagrange strain tensors \cite{FETM}. The $\eta$ for monolayer RuF$_4$ is:
\begin{equation}
\eta =\left[\begin{array}{cc}0.013 & 0 \\ 0 & -0.045\end{array}\right],
\end{equation}
which implies that there is $1.3\%$ tensile strain along the $a$-axis and $4.5\%$ compressive strain along the $b$-axis when comparing with the paraelastic state $P$. And the ferroelastic strain can be determined as $(|a|/|b|-1) \times 100\% $, which is $6.07\%$ for monolayer RuF$_4$. Compared with typical ferroelastic phosphorene analogues such as SnS ($4.9\%$) and SnSe ($2.1\%$) \cite{Wu2016}, RuF$_4$ could exhibit a more prominent ferroelastic switching signal in applications. Also, its large ferroelastic anisotropy can help the monolayer to be more flexible and suffer more elastic deformation \cite{liumin:science,YouPrb}.

Furthermore, it is worth noting that the in-plane spin orientation can be tuned accompanying the ferroelastic switching, leading to strong magnetoelasticity. As mentioned before, in the state $S$, the spin direction is mainly along the $a$-axis with a net magnetization along the $b$-axis. While in the equivalent state $S'$, the spin direction is mainly along the $b$ axis and the net magnetization points along the $a$-axis. Such a magnetic rotation process is illustrated in Fig.~\ref{fig5}(b). In the intermediate region ($a\approx b$),  the spin orientation (and magnetization) starts to change from the original ones ($a$ or $b$) to the diagonal one, as demonstrated in Fig.~S2 \cite{supp}.

\section{Conclusion}
In summary, our first-principles study has revealed the non-trivial physical properties of AFM monolayer RuF$_4$, which can be exfoliated from its bulk phase with small energy consumption and keeps dynamically stable. Its weak ferromagnetism, which originates from canting of AFM texture, is mainly dominated by the DM interaction, which can persist up to a moderate $T_{\rm N}=45.5$ K.

Its interesting physical properties include MOKE and magnetoelasticity. Remarkably, considerable L-MOKE occurs in this mainly AFM monolayer due to the $PT$ symmetry breaking, similar to those of regular ferromagnets. In addition, strong magnetoelasticity is found, with direct coupling among the spin orientation, magnetization, and ferroelastic distortion.

Our work opens a window to pursue more kinds of 2D magnets and multiferroics, beyond the plain ferromagnets based on those $3d$ transition metal compounds. Our theoretical work will stimulate more experimental studies on monolayer RuF$_4$ and related systems, towards both fundamental discoveries and potential spintronic applications.

\begin{acknowledgments}
 We thank Profs. Wanxiang Feng and Weiwei Lin for helpful discussions on MOKE. The work was supported by the National Natural Science Foundation of China (Grant Nos. 12104089  and 11834002), the Natural Science Foundation of Jiangsu Province (Grant No. BK20200345), and Postgraduate Research \& Practice Innovation Program of Jiangsu Province (Grant No. KYCX21\_0079). We thank the Big Data Center of Southeast University for providing the facility support on the numerical calculations.
\end{acknowledgments}

\bibliography{RuF4_ref}
\bibliographystyle{apsrev4-2}
\end{document}